\documentclass[a4paper]{article}

\usepackage[utf8]{inputenc}
\usepackage[T1]{fontenc}

\usepackage{amsfonts}
\usepackage{mathrsfs}

\usepackage{amssymb}
\usepackage{amsthm}
\usepackage{amsmath}

\usepackage{dsfont}

\emergencystretch=\maxdimen
\binoppenalty=\maxdimen
\relpenalty=\maxdimen

\usepackage{enumitem}

\usepackage{mathtools}

\newcommand{\ev}[1]{
	\mathbb{E}\left[ #1 \right]
}

\newcommand\R{
	\mathbb{R}
}

\newcommand{\var}{
	\text{var}
}

\DeclarePairedDelimiter\abs{\lvert}{\rvert}%
\DeclarePairedDelimiter\norm{\lVert}{\rVert}%

\makeatletter
\let\oldabs\abs
\def\abs{\@ifstar{\oldabs}{\oldabs*}}
\let\oldnorm\norm
\def\norm{\@ifstar{\oldnorm}{\oldnorm*}}
\makeatother

\usepackage{amsthm}
\usepackage[capitalise]{cleveref}

\newtheoremstyle{break}
  {\topsep}{\topsep}%
  {\upshape}{}%
  {\bfseries}{}%
  {\newline}{}%
\theoremstyle{break}
\newtheorem{thm}{Theorem}[section]

\newtheorem{prop}[thm]{Proposition}

\newtheorem{dfn}{Definition}[section]

\theoremstyle{remark}
\newtheorem{rmk}{Remark}

\theoremstyle{plain} %
\newcommand{\thistheoremname}{}
\newtheorem{genericthm}[thm]{\thistheoremname}

\newtheoremstyle{mystyle}%
  {}%
  {}%
  {\itshape}%
  {}%
  {\bfseries}%
  {.}%
  { }%
  {\thmname{#1}\thmnumber{ #2}\thmnote{ (#3)}}%

\crefalias{thm}{theorem}
\crefalias{lem}{lemma}
\crefalias{prop}{proposition}
\crefalias{cor}{corollary}
\crefalias{dfn}{definition}
\crefalias{conj}{conjecture}
\crefalias{exmp}{example}

\usepackage{graphicx}

\usepackage[automark]{scrlayer-scrpage}
\usepackage{titlesec}
\usepackage[toc]{appendix}
\usepackage[space]{grffile}
\usepackage{svg}

\def\UsePng{}

\clearpairofpagestyles
\ohead{\headmark}
\ofoot*{\pagemark}
\addtokomafont{pagenumber}{\footnotesize}

\let\oldpar\paragraph
\renewcommand{\paragraph}[1]{\oldpar{#1} \mbox{} \\}

\graphicspath{ {./graficos/} }

\providecommand{\keywords}[1]
{
  \small	
  \textbf{\textit{Keywords---}} #1
}

\usepackage{algorithm}
\usepackage[noend]{algpseudocode}
\usepackage{placeins}
\usepackage{caption}

\makeatletter
\renewcommand{\Function}[2]{%
  \csname ALG@cmd@\ALG@L @Function\endcsname{#1}{#2}%
  \def\jayden@currentfunction{#1}%
}
\newcommand{\funclabel}[1]{%
  \@bsphack
  \protected@write\@auxout{}{%
    \string\newlabel{#1}{{\jayden@currentfunction}{\thepage}}%
  }%
  \@esphack
}
\makeatother

\newcommand\X{
	\mathcal{X}
}
\newcommand\Y{
	\mathcal{Y}
}
\newcommand\LL{
	\mathcal{L}
}

\numberwithin{equation}{section}

\begin{document}

\author{Henrique Guerreiro \footnote{Supported by FCT Grant SFRH/BD/147161/2019.}
\\
hguerreiro@iseg.ulisboa.pt
 \and João Guerra 
\footnote{Partially supported by the project CEMAPRE/REM-UiDB/05069/2020 - financed by FCT/MCTES through national funds.} 
 \\
jguerra@iseg.ulisboa.pt\\
}
\date{
ISEG - School of Economics and Management, Universidade de Lisboa \\
REM - Research in Economics and Mathematics, CEMPARE \\
Rua do Quelhas 6, 1200-781 Lisboa, Portugal \\
\vspace{10pt}
\today 
}
\title{
Least squares Monte Carlo methods in stochastic Volterra rough volatility models
}

\bibliographystyle{apalike}

\maketitle
\begin{abstract}
In stochastic Volterra rough volatility models, the volatility follows a truncated Brownian semi-stationary process with stochastic vol-of-vol. Recently, efficient VIX pricing Monte Carlo methods have been proposed for the case where the vol-of-vol is Markovian and independent of the volatility.
Following recent empirical data, we discuss the VIX option pricing problem for a generalized framework of these models,  where the vol-of-vol may depend on the volatility and/or not be Markovian. In such a setting, the aforementioned Monte Carlo methods are not valid. Moreover, the classical least squares Monte Carlo faces exponentially increasing complexity with the number of grid time steps, whilst the nested Monte Carlo method requires a prohibitive number of simulations.
By exploring the infinite dimensional Markovian representation of these models, we device a scalable least squares Monte Carlo for VIX option pricing. We apply our method firstly under the independence assumption for benchmarks, and then to the generalized framework. We also discuss the rough vol-of-vol setting, where Markovianity of the vol-of-vol is not present.  We present simulations and benchmarks to establish the efficiency of our method.
\end{abstract}


\cleardoublepage

\section{Introduction}

Since the seminal papers \cite{VolIsRough} and \cite{PricingRough}, rough volatility modeling has enjoyed a tremendous focus of interest by the academic community. In such setting, the log-volatility behaves similarly to a fractional Brownian motion with Hurst parameter $H<1/2$. For this reason, the sample paths are rougher, in the sense that they exhibit a lower  H{\"o}lder continuity exponent, than those of standard Brownian motion. The rBergomi model, one of the first rough volatility models, adjusts very well to the SP500 index with a small number of parameters. Some of the relevant literature regarding rough volatility includes \cite{Microstructure}, which provides a micro-structural foundation for rough volatility; \cite{ShortTimeAlos} and \cite{ShortTimeAlosFuka}, which discuss asymptotic results; as well as a treatment of hedging strategies for the rough Heston model, in \cite{PerfectHedging}.

A major drawback of the rBergomi model is that it produces flat smiles for the VIX index options market. There are multiple ways to tackle this problem (see \cite{Alos18}). As in the Black-Scholes model volatility was made random to prevent a flat smile for the SP500, it may be natural to introduce stochastic vol-of-vol to produce non-flat VIX smiles. The stochastic Volterra models (SVM) of \cite{ModulatedVolterra} generalize the rBergomi model in a natural way, seeing it as a truncated Brownian semi-stationary process with constant intermittency. The variance process is
$$
v_u = A_0(u) \exp\left( 2 \int_0^u g_u(s) \sqrt{\Gamma_s} dW_s \right),
$$
for a standard Brownian motion (sBm) $W$, rough kernel $g$, vol-of-vol $\Gamma$ and a deterministic function $A_0$. When the vol-of-vol $\Gamma$ is made non-constant, it is possible to produce upward VIX smiles that resemble those observed in the market.

In the framework of \cite{ModulatedVolterra}, $\Gamma$ is assumed Markovian and independent of $W$, which lets the authors develop an efficient pricing mechanism for the VIX. There are however good reasons to pursue a more general framework where these assumptions do not hold. First, a correlation between price and vol-of-vol contributes to upward slopping VIX smiles (see \cite{Joint-VIX-rHeston}), which would make the models more flexible. Secondly, there is empirical evidence that variance and vol-of-vol are correlated, and that vol-of-vol may even also display the rough property (see \cite{Vol-of-vol-rough}). This more general framework could allow for a more flexible solution to the SP500-VIX joint calibration problem (see \cite{Joint-VIX-Discrete} and \cite{Joint-VIX-rHeston}).

The main challenge to the more general framework is the computational cost of VIX option pricing. Although Monte Carlo pricing under rough volatility has become more efficient (see \cite{Hybrid2017} and \cite{Turbo2018}), it is still computationally challenging. Moreover, the proposed Monte Carlo algorithms are very model dependent. To the best of our knowledge, no treatment of the general framework of the SVM has been performed, and no VIX pricing methods have been proposed.

In the absence of a model specific method to price VIX options, we may turn to more model-free methods. The nested Monte Carlo (NMC) method is always available but suffers from severe performance issues due to the extremely high number of simulations required. The classical least squares Monte Carlo (LSMC) method (see \cite{LSMCSchwartz}) can be used to price VIX derivatives for a wide range of models (see \cite{VIX-LSMC}) but unfortunately works poorly in a non-Markovian setting, as the state variables will be infinite dimensional. We will explore how the LSMC can be tailored to the SVM to work in this non-Markovian setting. Recently, in  \cite{ChaosLSMC}, while following a very different approach from ours, the authors propose a variant of the LSMC for Bermudan option pricing that works under non-Markovian conditions. They achieve this by replacing the classical regression by a Wiener chaos expansion.

In this paper, by exploring the structure of the infinite dimensional state variable, we develop a LSMC method for pricing VIX options for the general framework of the SVM. We fully detail the algorithm and present numerical evidence to establish its performance, using the standard NMC as benchmark. We conclude our LSMC is able to accurately price VIX options, achieving the same degree of precision as the NMC at a fraction of the computational cost.

The paper is organized as follows. In \cref{sec:VIX}, we will introduce the VIX pricing problem and explain how the NMC and LSMC methods can be used to tackle it. We will draw attention to the challenges to these methods in a non-Markovian setting. 
In \cref{sec:SVM}, we shall then formulate the SVM and discuss its main properties. Namely, we will discuss how to surpass the main obstacles to efficient VIX pricing in the more restricted framework. Afterwards, we will present our LSMC pricing method and show how it can be applied in a more general framework that allows for dependence of vol-of-vol and volatility. Then, in \cref{sec:rInter}, we briefly explore the rough vol-of-vol model and discuss the main challenges of the LSMC in that setting. In \cref{sec:num}, we explore some numerical simulations and compare the results with benchmarks in order to access the performance of the proposed LSMC method. Finally, in \cref{sec:conc}, we present some concluding remarks and suggest further related research problems.

\section{VIX Pricing}
\label{sec:VIX}

\subsection{The VIX index}

Let $v$ denote the variance and $\left(\mathcal{F}_t\right)_{t \geq 0}$ be the filtration generated by the processes which drive $v$. To simplify, we will work under the assumption of zero interest rates. The CBOE VIX index (see \cite{VIX-WhitePaper}), which reflects the market's expectations on future (30 day) volatility, can be defined by the expression
\begin{equation}
VIX_T = \sqrt{\frac{1}{\Delta} \ev{  \int_T^{T+\Delta} v_u \, du \mid \mathcal{F}_T }},
\end{equation}
where $T>0$ is a fixed time and $\Delta=30$ days is the VIX horizon.
Let 
\begin{equation}
\xi_t(u) = \ev{ v_u \mid \mathcal{F}_t }
\end{equation}
be the forward variance curve.
By a simple application of Fubini's theorem, we can write the VIX in terms of the forward variance curve
\begin{equation}
VIX_T  = \sqrt{\frac{1}{\Delta} \int_T^{T+\Delta} \xi_T(u) \, du}.
\label{eq:vix}
\end{equation}
Options and futures can be traded on the VIX, making it a key risk management instrument. If we are to model volatility, we must be able to price VIX options and futures with a reasonable computational cost.
\subsection{Nested Monte Carlo}
Since the VIX is essentially the square-root of a conditional expectation, there is no direct way to price its options or futures without evaluating the conditional expectation itself. The most general method available for this purpose is the NMC. 

In general, let us denote by $\theta$ the parameters of a model (e.g. $H$ in the rBergomi model) and by $\beta$ the initial state variables (e.g. the initial variance in the rBergomi model). Let us assume we possess a procedure \textit{ \ref{alg:sim} } that lets us simulate the model's state variables up a time horizon $T$, given the parameters and the initial state variables.  Let us also make the obvious assumption that the variance between $T$ and $T+\Delta$, and hence the integral $\int_T^{T+\Delta} v_u \, du$, can be obtained from the state variables between times $T$ and $T+\Delta$. Our goal is to create a simulation of $N$ values of 
\begin{equation}
Y = \ev{  \int_T^{T+\Delta} v_u \, du \mid \mathcal{F}_T }
\end{equation}
since the VIX can be trivially obtained from it. We would then have a traditional Monte Carlo simulation of the VIX and be able to compute the payoff of any contingent claim. 

Notice that in a non-Markovian setting (as in rough volatility models), we must resort to infinite dimensional state variables. In the rBergomi model, this is simply the forward variance curve. In practice, this problem is partially circumvented by projecting such variables into finite dimensional vector spaces. 

Bellow we can find a description of the NMC algorithm.
\begin{algorithm}
\caption{Nested Monte Carlo}\label{alg:nmc}
\begin{algorithmic}[1]
\Function{Simulate}{$\theta,\beta, t$} \funclabel{alg:sim} \ \Comment Provided by the model
\EndFunction 
\Function{NestedMonteCarlo}{$\theta,\beta, T, N, M$}
\For {$i=1$ to $N$}\Comment{Outer paths}
\State $\beta'_i \gets $ \Call{Simulate}{$\theta,\beta, T$} \ \Comment Simulates outer path
\For {$j=1$ to $M$} \Comment{Inner paths}
\State $\beta_j^{(i)}\gets \Call{Simulate}{\theta,\beta'_i, \Delta}$ \ \Comment Simulates inner path
\State $ X_j^{(i)} \gets \Call{InnerX}{\beta_j^{(i)}}$ \ \Comment Inner estimate
\EndFor
\State $Y_i \gets \Call{Mean}{  X^{(i)}   }$
\ \Comment Averages inner estimates
\EndFor
\State \textbf{return} $\beta', Y$ \Comment{State variables and target variable}
\EndFunction 
\end{algorithmic}
\end{algorithm}
\subsection{Least squares Monte Carlo}
As it can be readily noticed by observing the algorithm, the NMC method is extremely slow. This is of course due to the fact that for each outer path, it needs to perform a further Monte Carlo simulation. As a consequence, this algorithm is often impractical. The LSMC method provides a more efficient alternative by attempting to reduce the number of simulations. The algorithm produces $N$ intermediate (potentially noisy) approximations of the target, which are then fed into a regression model. Intuitively, the regression model averages out the noise across the intermediate simulations and is able to learn the true function. This regression model is then used to produce a good estimate of the target with a lower cost. Depending on the variant, the regression model may or may not use the same data for both fitting and predicting. Moreover, when $K$ outer paths are generated but only $N$ are used for fitting, in the last step of the algorithm we may either use all the $K$ outer paths generated for prediction, or only the $K-N$ that were not used to fit the regression.
\begin{algorithm}
\caption{Least squares Monte Carlo}\label{alg:lsmc}
\begin{algorithmic}[1]
\Function{Simulate}{$\theta,\beta, t$} \ \Comment Provided by the model
\EndFunction 
\Function{RegressionModel}{$X,Y$} \funclabel{alg:reg} \Comment Fits a certain  regression model
\EndFunction
\Function{LeastSquaresMonteCarlo}{$\theta,\beta, T, N, M, K$}
\For {$i=1$ to $K$} \Comment Generate $K$ outer paths
\State $\beta''_i \gets $ \Call{Simulate}{$\theta,\beta, T$} 
\EndFor
\State $\{i_1, i_2, ..., i_N\} \gets \Call{Sample}{K, N}$  \Comment{Sample $N \leq K$ paths}
\For {$k=1$ to $N$}\Comment{Outer paths}
	\For {$j=1$ to $M$} \Comment{Inner paths}
	\State $\beta_j^{(i_k)}\gets \Call{Simulate}{\theta,\beta'_{i_k}, \Delta}$ \ \Comment Simulates inner path
	\State $ X_j^{(i_k)} \gets \Call{InnerX}{\beta_j^{(i_k)}}$ \ 		\Comment Inner estimate
	\EndFor
	\State $Y'_{i_k} \gets \Call{Mean}{ X^{(i_k)} }$ \Comment Averages inner estimates
\EndFor
\State $R \gets \Call{RegressionModel}{\beta',Y'}$
\State $Y \gets R(\beta'')$ \Comment Regression prediction
\State \textbf{return} $Y$
\EndFunction 

\end{algorithmic}
\end{algorithm}

\subsubsection*{Increasing LSMC performance}
\label{rmk:lsmc}
Note that only the $N$ paths used to fit the regression require inner simulations. Thus, one of the main advantages of the LSMC is that there is great flexibility in managing the computational budget, which can be seen as the total number of simulations $K + N \times M$. To start, one might simply choose $K=N$ and see the LSMC as a regularization procedure on top of the NMC, in the sense that a lower number of inner simulations is needed to arrive at the same accuracy. Nevertheless, one of the great advantages of the LSMC is that once the regression model is fitted, it can be used to make fast predictions for unseen values of the predictors. This is especially relevant in our context of option pricing, since out-of-the-money options may require a considerable number of paths $K$ to be evaluated. The cost of such evaluation would be prohibitive using a NMC, where $N=K$. However, provided the regression model is of good quality, it can be used to efficiently generate $K$ paths of the target variable, even if only $N << K$ paths were used to learn the mapping. In fact, we can even use the information about the distribution of the predictor contained in the $K$ outer paths to inform the choice of the $N$ paths used for regression fitting. 

The choice of $N, K$ and $M$ depends on the nature of the problem and the choice of regression model. For instance, if the regression model is very sensitive to overfitting, we may need to choose a higher value of $M$ and even of $N$. The complexity of the regression model also plays a role: a simpler regression model may not need to see as many values of the target function, and thus may require a smaller $N$, allowing for higher $M$.
Note that if the functional form of the target function is known, the LSMC can potentially learn it very efficiently since only a small $N$ might be needed. Finally, as the dimension of the predictor gets higher, so will $N$, especially for more complex regression models.

\subsection{Challenges of non-Markovian modeling}

The regression model presented above is used to approximate a certain random variable $Y$, using the state variables $\beta \in \R^m$ as predictors. Since we are trying to approximate an arbitrary function $f$, a classical linear regression model will be inadequate if the function is highly non-linear. A natural strategy is to approximate the projection of $f$ into a finite dimensional vector space. As the Hermite polynomials are an orthogonal basis of $L^2$, we can write
\begin{equation}
f = \tilde{f}+ \varepsilon = \sum_{k=1}^n \alpha_k H_k + \varepsilon,
\end{equation}
where $(H_k)_{k \leq n}$, are the $n$ Hermite polynomials with degree no greater than $d$ in $\R^m$, and $\varepsilon$ is an error term. We may find $\alpha_k$ using classical linear regression, with predictors $H_k(\beta_i)_{k=1}^n, i=1,..., N$. The price we have to pay, however, is the exponentially increasing number of predictors with respect to the number of state variables $m$. 

In Markovian models, it is easy to encode all information known at time $T$ in a vector with only a few entries. In non-Markovian models, however, because we have to project the infinite dimensional state variable of the process into a finite dimensional vector space, we will have a vector $\beta_i$ with tens, or even hundreds, of entries. This challenge turns the ``classical'' LSMC as described above, unfeasible. 

In the next section, we will explore a very general family of models, of which the rBergomi model is a special case. The LSMC can be adapted to become a valid alternative for VIX option pricing in these models when some of the ``nicer'' analytical conditions are removed.

\section{Stochastic Volterra Models}
\label{sec:SVM}

The stochastic Volterra models of \cite{ModulatedVolterra} provide a promising framework to tackle the SP500/VIX Joint Calibration Problem.
In these models, the variance behaves as a truncated Brownian semi-stationary process (TBSS). They generalize the rBergomi model and are able to produce an upward slopping VIX smile. We start with the definition of the truncated Brownian semi-stationary process (TBSS), following \cite{SemiStationaryFirst} and \cite{Hybrid2017}.

\begin{dfn} 
We say $X$ is a truncated Brownian semi-stationary process if
\begin{equation}
X_u = \int_0^u \sqrt{\Gamma(s)} g(u-s)dW_s,
\end{equation}
for a sBm $W$, a predictable locally bounded process $\Gamma$, and a Borel measurable function $g$. The processes $\Gamma$ and $W$ may possibly be dependent.
\end{dfn}
In order to simulate paths of a TBSS, we may use the Hybrid scheme, introduced in \cite{Hybrid2017}, which is able to deal with the singularity posed by rough kernels. The main idea of the method is to approximate the integral near the singularity by that of a power function, while using regular Riemann sums away from the singularity. The final approximation then involves Wiener integrals of a power function as well as a Riemann sum, and hence the name \textit{hybrid} scheme. In a SVM, the log-variance is modeled by a TBSS:
\begin{dfn}[Stochastic Volterra model]
Let $B$ and $W$ be two $\rho$-correlated Brownian Motions, with $-1 < \rho < 1$. We say that $(S, v)$ follows a stochastic Volterra model if
\begin{equation}
d S_t = S_t \sqrt{v_t} dB_t,
\end{equation}
where the variance process is given by
\begin{equation}
v_u = A_0(u) e^{2X_u}
\end{equation}
and $X_u$ is a TBSS given by
\begin{equation}
X_u = \int_0^u \sqrt{\Gamma_s} g_u(s) \, ds.
\end{equation}
The rough kernel $g$ is in our setting given by
\begin{equation}
g_u(s) = (u-s)^{H-1/2}
\end{equation}
for Hurst parameter $H$, with $0<H<1/2$ and a deterministic function $A_0$. The vol-of-vol process $\Gamma$ starts at a given constant $\gamma >0$.
\end{dfn}

\begin{rmk}
In the paper \cite{ModulatedVolterra}, the vol-of-vol process is assumed to be independent from $W$ and Markovian. In the definition above we are intentionally dropping those assumptions. This leads to a more general framework, with the obvious cost of analytical and computational difficulties. Attempting to overcome those difficulties is the main subject of our paper.
\end{rmk}

The rBergomi model of \cite{PricingRough} can be seen as a particular case of the SVM, one in which the vol-of-vol is constant and given by
\begin{equation}
\label{eq:ConstBerg}
\Gamma(s) \equiv \gamma = \frac{\eta^2 H}{2},
\end{equation}
for a fixed $\eta > 0$.

The rBergomi model is able to calibrate the SP500 very well but cannot calibrate the VIX smile since the market exhibits upward slopping smiles whilst the rBergomi model, by virtue of its constant vol-of-vol, produces approximately flat VIX smiles. The extra randomness introduced by stochastic vol-of-vol provides the flexibility needed for an upward slopping VIX smile, as it was showed in \cite{ModulatedVolterra}. 

In order to develop a successful VIX option pricing method in the general framework, let us try to find an infinite dimensional representation for the SVM. 

\begin{prop}
\label{prp:infdim}
For $u \geq t > 0$, let
\begin{equation}
A_t(u) = \frac{\xi_t(u)}{h_t(u)},
\end{equation}
where 
\begin{equation}
\label{eq:hT}
h_t(u) = \ev{ E_{t, u}(u) \mid \mathcal{F}_t }
\end{equation}
and
\begin{equation}
E_{p, q}(u) = \exp\left( 2 \int_p^q g_u(s) \sqrt{ \Gamma_s } dW_s\right).
\end{equation}
We have the time-invariant decomposition
$$
v_u = A_0(u) E_{0, u}(u) = A_t(u) E_{t, u}(u) \, \forall \, u \geq  t > 0.
$$
In addition, if $\Gamma$ is Markovian, the process $A_t$ provides an infinite dimensional Markovian representation of the SVM in the sense that conditional on $A_t$ and $\Gamma_t$, $v_u$ is independent of $\mathcal{F}_t$.
\end{prop}

The (simple) proof of this fact can be found in \cref{App1}.

\begin{rmk} \label{rmk:rB-hT}
In the rBergomi model, $\Gamma \equiv \gamma$, constant. In such setting
\begin{align*}
h_t(u) &= \ev{
 \exp \left(2 \sqrt{\gamma}
  \int_t^u g_u(s)  d W_s 
\right) \mid \mathcal{F}_t } \nonumber \\
&= \exp\left(
2 \gamma
\int_t^u (u-s)^{2H-1}
\right) \nonumber \\
&= \exp \left(
\frac{1}{2} \eta^2 (u-t)^{2H} 
\right) ,
\end{align*}
where we have used \cref{eq:ConstBerg}. This means $A_t(u)$ and $\xi_t(u)$ are equivalent in the rBergomi model, whereas in the SVM we need to compute the (random) conditional expectation $h_t(u)$ in order to relate them.
\end{rmk}
\begin{rmk} Note that regardless of whether $\Gamma$ is Markovian, the ``hard'' part of VIX pricing is the curve $h_T$. Indeed,
$$
\Delta VIX_T^2 = \int_T^{T+\Delta} \xi_T(u) \, du =  \int_T^{T+\Delta}  \frac{ \xi_0(u) }{h_0(u) } E_{0, T}(u) h_T(u)  \, du.
$$
The value $E_{0, T}(u)$ acts as a state variable and can be computed simply using Riemann sums. The initial forward variance curve $\xi_0$ can be observed from the market or assumed flat and taken as a parameter. As for $h_0$, it can be written as follows:
$$
h_0(u) = \ev{ E_{0, u}(u) } = \ev{ E_{0, T}(u)  \ev{E_{T, u}(u) \mid \mathcal{F}_T } } = \ev{  E_{0, T}(u) h_T(u) }.
$$
Thus, since we know $E_{0, T}(u)$ and $h_T(u)$, we may estimate $h_0(u)$ by averaging the product of these two variables. Alternatively, $h_0$ can be estimated through a (simple) Monte Carlo simulation.
\end{rmk}

\subsection{Independent and Markovian vol-of-vol}
\label{sec-IndMark}

In the discussion of the Markovian and independent vol-of-vol process, we follow \cite{ModulatedVolterra}. Let us assume $\Gamma$ is Markovian and independent of the sBm $W$. Then, for any $u \geq t > 0$, 
\begin{equation}
h_t(u) = \ev{ E_{t, u}(u) \mid \Gamma_t }.
\end{equation}
By conditioning on the full path of $\Gamma$ between $t$ and $u$, $E_{t, u}(u)$ is conditionally log-normal with mean zero, so that
\begin{equation}
h_t(u) = \ev{ \exp\left( 2 \int_t^u g_u^2(s) \Gamma(s) \, ds \right) \mid \Gamma_t }.
\end{equation}
If $\Gamma$ is assumed to be exponentially affine, along some integrability conditions, the above conditional expectation can be written in terms of the solution to a certain ODE and the values $\Gamma_0$ and $\Gamma_T$. For details, we refer to \cref{App1}. This leads to an efficient framework for VIX pricing since there is no need for inner simulations in order to compute the conditional expectation. An example of a process that satisfies the above conditions, as provided in \cite{ModulatedVolterra}, is the CIR process:
\begin{equation}
d \Gamma_t = \theta(m - \Gamma_t) \, dt + \delta \sqrt{\Gamma}d Z_t,
\end{equation}
where $\theta, m, \delta >0$ and $Z$ is a sBm independent of $W$.

\subsection{Dropping independence}

There is some evidence that dropping one (or both) the assumptions of independence or Markovianity is justified. First, the vol-of-vol process might be itself a rough volatility process, which suggests a non-Markovian vol-of-vol (see \cite{Vol-of-vol-rough}). Additionally, vol-of-vol is higher in periods of crisis, suggesting a negative correlation with the asset price, and this contributes to upward slopping VIX smiles (see \cite{Joint-VIX-rHeston} and \cite{NotInd}). Even if we only wished to correlate vol-of-vol and price, this might require variance and vol-of-vol to be correlated. In order to see this, let us take the CIR example above and assume that $B, W$ and $Z$, which are the standard Brownian motions driving the price, volatility and vol-of-vol, respectively, form a 3-dimensional Brownian motion with correlation matrix 
$$
\Sigma = \begin{bmatrix}
1 & \rho & \rho_S \\
\rho & 1  & \rho_V \\
\rho_S & \rho_V & 1 \\
\end{bmatrix}.
$$
The determinant of this matrix is of course
$$
1 - \rho^2 - \rho_S^2  - \rho_V^2 + 2\rho \rho_S \rho_V.
$$
If variance and vol-of-vol are independent, $\rho_V=0$ and the determinant becomes
$$
1 - \rho^2 - \rho_S^2.
$$
But since we know from empirical data that $\rho \approx -1$, for the matrix to be positive definite we must have $\rho_S \approx 0$.

The question now is: can we generalize the above framework by dropping one of the two crucial assumptions (independence and Markovianity) whilst keeping the VIX pricing methods with a reasonable running time?

\subsection{Least squares Monte Carlo}

Let us drop the assumption that $\Gamma$ and $W$ are independent but assume $\Gamma$ is still Markovian. Let us take the infinite dimensional state variable $h_T$. Then, we fix a time grid  $(u_j)_{j=1}^n$ between $T$ and $T+\Delta$ and obtain the projection  $Y = (h_T(u_j))_{j=1}^n$. We will use the LSMC method to approximate $Y$. Since $\Gamma$ is Markovian, we have
\begin{equation}
Y_j = h_T(u_j) = \ev{ 2 \int_T^{u_j} \sqrt{\Gamma_s} g_{u_j}(s) \mid \Gamma_T } = f_j(\Gamma_T),
\end{equation}
for some deterministic function $f_j$. Thus, we may write this as a classical multivariate regression problem:
\begin{equation}
Y = f(\Gamma_T) = \tilde{f}(\Gamma_T) + \varepsilon,
\end{equation}
where $\tilde{f}$ is the deterministic approximation function and $\varepsilon$ is the stochastic error term. Note that the number of predictors is simply one (the vol-of-vol process at time $T$). Increasing the number of grid points will only increase the size of the output vector, leading to a linear, and not exponential, increase in complexity.

In order to take maximum advantage of the LSMC, especially in this one-dimensional problem, we allow for $N < K$ (see \cref{rmk:lsmc}). In order to use the information contained in the $K$ outer paths, we sample $N$ out of the $K$ generated in order to fit the regression model. 

Fix a time grid $(u_j)_{j=1}^n$ between $T$ and $T+\Delta$. The LSMC method for SVM is then as follows. 

\begin{enumerate}
\item Run an outer simulation and obtain $K$ realizations of the state variables $\Gamma_T$ and $[E_{0,T}(u_j)]_{j=1}^n$.
\item Of these $K$ outer paths, take a sample of $N \leq K$.
\item For each outer value $\Gamma_T^{(i)}$ of vol-of-vol, run an inner simulation to obtain $[E_{T, u_j}^{(i)}(u_j)]_{j=1}^n$, which is a TBSS with initial vol-of-vol $\Gamma_T^{(i)}$. 
\item Compute the mean over inner simulations and obtain intermediate estimates $y_j'$ of $h_T(u_j)$ for  $j=1,...,n$.
\item Fit the regression model with predictor $\Gamma_T$ and dependent vector $(y_j')_{j=1}^n$.
\item Use the fitted regression model to obtain the final estimates of $[h_T(u_j)]_{j=1}^n$ by evaluating the $K$ initially generated outer paths.
\item Obtain an estimate for $h_0(u_j)$ by averaging the product $E_{0, T}(u_j) h_T(u_j)$, for each $j=1,...,n$.
\item Compute the forward variance curve using the formula
\begin{equation}
\xi_T(u_j) =\frac{ \xi_0(u_j) }{h_0(u_j) } E_{0, T}(u_j) h_T(u_j), j=1, ..., n.
\end{equation}
\item Compute the VIX value by approximating the integral in \cref{eq:vix} using a classical quadrature method.
\end{enumerate}

\begin{rmk}
We note that the choice of regression model need not be the linear regression on a function basis like the Hermite polynomials. Indeed, non-linear non-parametric methods such as Random Forests and Neural Networks can be used. For more details about Random Forests, we refer to \cref{App2}.
\end{rmk}

\section{Rough vol-of-vol}
\label{sec:rInter}

In \cite{Vol-of-vol-rough}, the authors have provided empirical evidence that volatility of volatility is also rough. They took high frequency data for the VIX index and computed its volatility, which acts as volatility of volatility. Then they showed its logarithm resembles a fractional Brownian motion with Hurst parameter less than $1/2$, similarly to what was done in \cite{VolIsRough} to establish roughness of volatility. Moreover, they estimated a positive correlation between volatility and volatility of volatility.

The (extended) SVM framework can accommodate for a rough vol-of-vol. Indeed, we may simply consider the vol-of-vol $\Gamma$ as a rough log-TBSS with constant volatility
\begin{align}
&\Gamma_s = Q_0(s) I_{0, s}(s), \\
&I_{0, s}(s) := \exp\left(
2 \sqrt{\nu} \int_0^s  g_s(t)  \, dZ_t
\right),
\end{align}
where $Z$ is a sBm correlated with $W$ and $B$, $\nu>0$ is a constant and $Q_0$ is a deterministic function. Since $\Gamma$ is a log-TBSS with constant volatility, using \cref{prp:infdim} and proceeding as in \cref{rmk:rB-hT}, we know the law of $\Gamma$ only depends  on $\mathcal{F}_T$  through 
\begin{equation}
\zeta_T(u) := \ev{ \Gamma_u \mid \mathcal{F}_T}.
\end{equation}
Let us then assume that the initial curve is flat $\zeta_0(u) \equiv \zeta_0$, where $\zeta_0$ is treated as a model parameter to be chosen. Then, as in the rBergomi model, we can obtain $\zeta_T$ by the formula
\begin{equation} \label{eqn:update-zeta}
\zeta_T(u) = q_T(u) \zeta_0(u) I_{0, T}(u),
\end{equation}
where $q_T$ is an analytically known deterministic function.
In this model, the vol-of-vol $\Gamma$ is neither Markovian nor independent of $W$. It therefore poses a formidable computational task. As seen above, in order to compute a VIX path we need the curve 
\begin{align}
h_T(u) &= \ev{ \exp\left( 2 \int_T^u g_u(s) \sqrt{\Gamma_s} \, dW_s \right) \mid \mathcal{F}_T } \\
&= F( \zeta_T(s)_{T \leq s \leq u} ),
\end{align}
since $\Gamma$ only depends on $\mathcal{F}_T$ through $\zeta_T$.

If we wish to run a simulation of this model and the relevant state variables we may proceed as follows:
\begin{enumerate}
\item Produce paths for the three dimensional Brownian Motion $(B, W, Z)$, where the covariance matrix is given.
\item Using the paths of $Z$, apply the hybrid scheme method (note constant volatility of the TBSS) to obtain paths for $\Gamma$ and $I_{0, T}(t)$ for $T \leq t \leq u$ with $t$ in some time grid. 
\item Apply the updating formula (\ref{eqn:update-zeta}) and obtain
$$
\zeta_T(t) =\zeta_0(t)  q_T(t)  I_{0, T}(t) =\zeta_0   q_T(t) I_{0, T}(t).
$$
\item For each obtained path of $\Gamma$, apply the hybrid scheme to compute the integral $\int_0^u g_u(s) \sqrt{\Gamma_s} \, dW_s$ and then proceed as usual.
\end{enumerate}

The LSMC method for rough vol-of-vol could then be applied as follows:
\begin{enumerate}
\item Run an outer simulation and obtain $K$ realizations of the state variables $\zeta_T(u_j)$ and $E_{0,T}(u_j)$ for $j=1,..., n$.
\item For each $i$ of $N \leq K$ selected outer paths, apply the procedure above to obtain $M$ inner paths of $[E_{T, u_j}^{(i)}(u_j)]_{j=1}^n$, which is a TBSS with rough volatility, where the (projection of the) initial curve is $[ \zeta^{(i)}_T(u_j) ]_{j=1}^n$. Again, we may choose $N < K$ if needed.
\item Compute the mean over inner simulations and obtain intermediate estimates $y_j'$ of $h_T(u_j)$ for $j=1,...,n$.
\item Fit the multivariate regression model with predictors $[\zeta_T(u_j)]_{j=1}^n$ and dependent vector $(y_j')_{j=1}^n$.
\end{enumerate}

The remaining steps are as in the Markovian vol-of-vol case. The added difficulty now is that the trick we used to escape an infinite dimensional predictor does not work, since even $h_T$ depends on $\mathcal{F}_T$ through an infinite dimensional object. Thus, the size of our predictor makes the usual linear regression on Hermite polynomials unpractical. In order to price VIX options in this setting one needs to consider the whole path $\zeta_T$ as a predictor. 

Although a full non-Markovian method for VIX option pricing is outside the scope of this article, we would like to point out some promising approaches. Since the structure of the dependence between $h_T$ and $\zeta_T$ is unknown, general high dimensional machine learning methods stand out as candidates for regression models in the LSMC. Of these we may name Random Forests, which are quite flexible and simple to implement, and Neural Networks, which are a focus of research and have been used before in the context of rough volatility (see \cite{DeepStone}, \cite{DeepBlanka} or \cite{DeepVol}). Finally, we would like to point to the Wiener Chaos approach of \cite{ChaosLSMC}, which can in principle be extended to rough volatility models.

\section{Numerical Experiments}
\label{sec:num}

The NMC for VIX pricing has three main sources of error: the number of time steps used for the discretization grid, the number of outer paths and the number of inner paths. In the context of LSMC, an extra source of error -- the regression model -- has to be considered. For each source, the error can be reduced by paying the corresponding computational price. Our main finding is that the LSMC allows to obtain accurate estimates at a reasonable computing cost, while the NMC requires a much higher computational cost in order to obtain the same accuracy. In fact, in some models, accurate pricing by NMC would require such a high computational cost as to be made impractical.

As seen in \cref{alg:lsmc}, we use only $N < K$ outer paths in order to fit our regression model but then use $K$ outer paths for the VIX by using the already fitted regression model. This key aspect of the LSMC allows it to vastly outperform the NMC. 

Thus, in order to access the quality of our LSMC, we proceeded as follows. First, we produced $K=50,000$ (outer) simulations of the model's state variables given the initial conditions and the parameters. Then, for various configurations of $N$ and $M$, we randomly select $N \leq K$ outer paths and produce $M$ inner simulations for these $N$ outer paths, which we use to fit various LSMC models for different regression model choices. The choice of $N$ and $M$ might depend on the regression model choice and the specification of the dynamics of $\Gamma$. For the NMC obviously $N=K$.

In line with LSMC practices, we consider the linear regression of Hermite polynomials as a regression model choice. The maximum degree of Hermite polynomials was set to $3$. Also, for their generality and wide applicability in machine learning, we also use Random Forests and Neural Networks.
In order to avoid overfitting, the tree maximum depth was set to $5$. All the remaining Random Forest hyper-parameters were set to the defaults of the \textit{RandomForestRegressor} class of the \textit{sklearn.enseble} Python library. For the Neural Network, we used three hidden layers of size 32, with a batch size of $32$ and a learning rate of $10^{-3}$. The sample was divided in training in validation, with $80\%$ used for training and $20\%$ for validation. Finally, we incorporate a simple Linear Regression in our analysis since $h_T$ is log-linear in the setting of \cref{sec-IndMark}. This simple toy example will also show us the power of the LSMC when the functional form the target function is known. The parameters of the model were chosen based partly on known best practices, as well  as some numerical tests, and for this reason they may be changed if the underlying model dynamics are changed. A further hyper-parameter optimization routine, based per example in cross validation, can be used to find the adequate hyper-parameters if one wishes to further increase the performance of LSMC.

In very general models, especially Random Forests but also Neural Networks, it is harder to generalize to tail values of the input. Thus, when sampling $N$ outer paths out of the initially $K$ generated outer paths, we extract a stratified sample. We split the input data in evenly spaced sub-intervals (strata) and sample from each one until a certain minimum sample goal is achieved or all points of the strata are sampled. The rest of the points are sampled uniformly to preserve the distribution of $\Gamma_T$. Thus, we make sure regions with higher probability mass have more weight while at the same time ensuring the regression model can cover the whole sample space.

The time horizon considered is $T=7$ days, where the VIX horizon is, as usual, $\Delta=30$ days. The number of time points per day is set to $n_d=7$, so that it makes only a small contribution to the  error. The initial forward variance curve is assumed flat $\xi_0(u) \equiv v_0$ and we set $v_0 = 0.013$. The Hurst index is set to be $H=0.1$. The correlation between $B$ and $W$ is $\rho = -0.95$ and the initial vol-of-vol is $\gamma = 0.05$. The vol-of-vol follows the CIR model where the parameters are chosen as follows: $\theta=0.4, \delta=0.8, \kappa = 0.8125$. We choose these parameters so that they both satisfy the conditions in \cite{ModulatedVolterra}, to allow for benchmarks, while also producing reasonable VIX smiles. The time grid for $[T, T+\Delta]$ is as in the Hybrid Scheme with $\kappa=1$:
$$
u_j = T + \frac{j}{365n_d}.
$$
All numerical simulations are performed on a Linux machine (Ubuntu 20.4), with AMD Ryzen 7 3800X CPU with 16 threads. The Python language was used, resorting to \textit{numpy} whenever possible for C-like speed. In order to train regression models, we used the \textit{scikit-learn} and \textit{pytorch} libraries. As most numerical tasks involved are \textit{embarrassingly parallel}, parallel computing was used to reduce running times.

\subsection{Markovian and independent vol-of-vol}

Let us first consider the case where the vol-of-vol $\Gamma$ is independent of $W$ and Markovian. In this case, we may use the approach of \cite{ModulatedVolterra} to know the true value of $h_T \mid \Gamma_T = \Gamma_T^{(i)}$ for each outer simulation $i$. This allows us to compute useful metrics for the LSMC method. Moreover, since the independence assumption is not used by the LSMC, we are lead to believe the quality of the algorithm generalizes well to the dependent case.

\subsubsection{Exploring the data}
\begin{figure}
\centering
\includegraphics[width=345 pt]{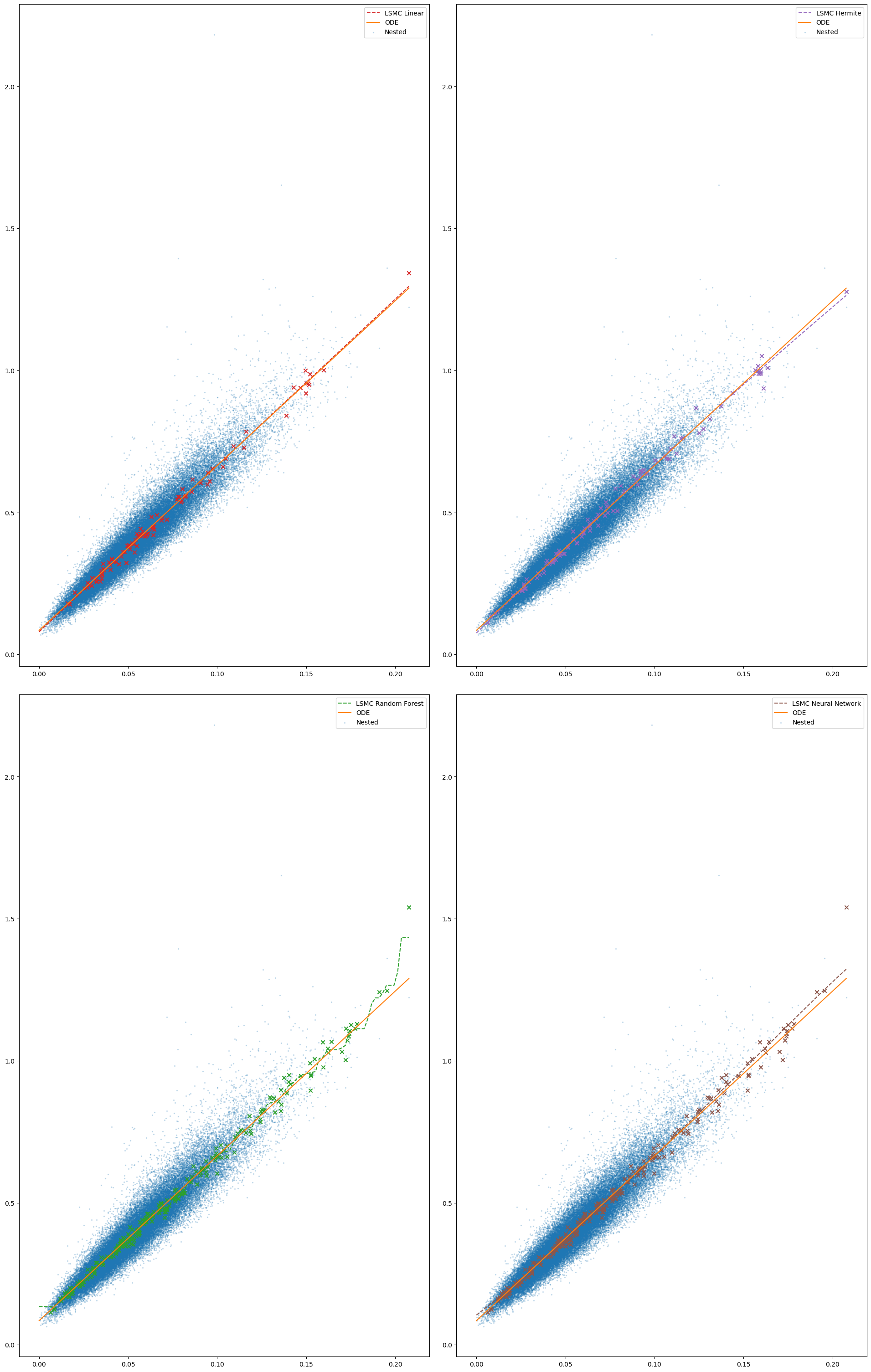}
\caption{Estimates of $h_T$ for various LSMC variants}
\label{hT-compare-ode}
\end{figure}

In order to start the data analysis, for a fixed $u$, we provide a scatter plot of the estimates for $h_T(u)$ produced by the NMC and the different variants of the LSMC algorithm in $log$ scale. The degree of noise is larger for larger $u$. We choose $u=u_{141}$, about two thirds of the way between $T$ and $T+\Delta$. In \cref{hT-compare-ode}, we can compare how the various regression models adjust to the data, while also getting a feel of how noisy the NMC data is with the maximum number of inner paths considered of $M=1000$. The proxy for the true value of $h_T$, coming from the ODE method described in \cref{App1}, is displayed for comparison in a solid line. The NMC, as expected, displays a great amount of noise.

In the independent Markovian case we know the function $h_T$ exactly:
\begin{equation}
h_T(u) = \exp( \psi(u-T) \Gamma_T + \phi(u-t) ),
\end{equation}
where $\psi$ and $\phi$ are as in \cref{thm:ModVol}. The function is log-linear and thus it is no surprise both the linear model and the Hermite present very good fits. Both the Random Forest and Neural Network present satisfactory fits. We will test these methods with a non-linear function in \cref{subsec:non-linear}.

\subsubsection{Variation of the number of inner paths}
We now wish to know the behavior of each algorithm as the number of paths increases. Since the configuration of $N$ and $M$ is different for each LSMC variant and NMC, and different regression models have different computing times, we found best to use running time to compare the performance of the models.

\begin{figure}
\centering
\ifdefined\UsePng
  \includegraphics[width=345 pt]{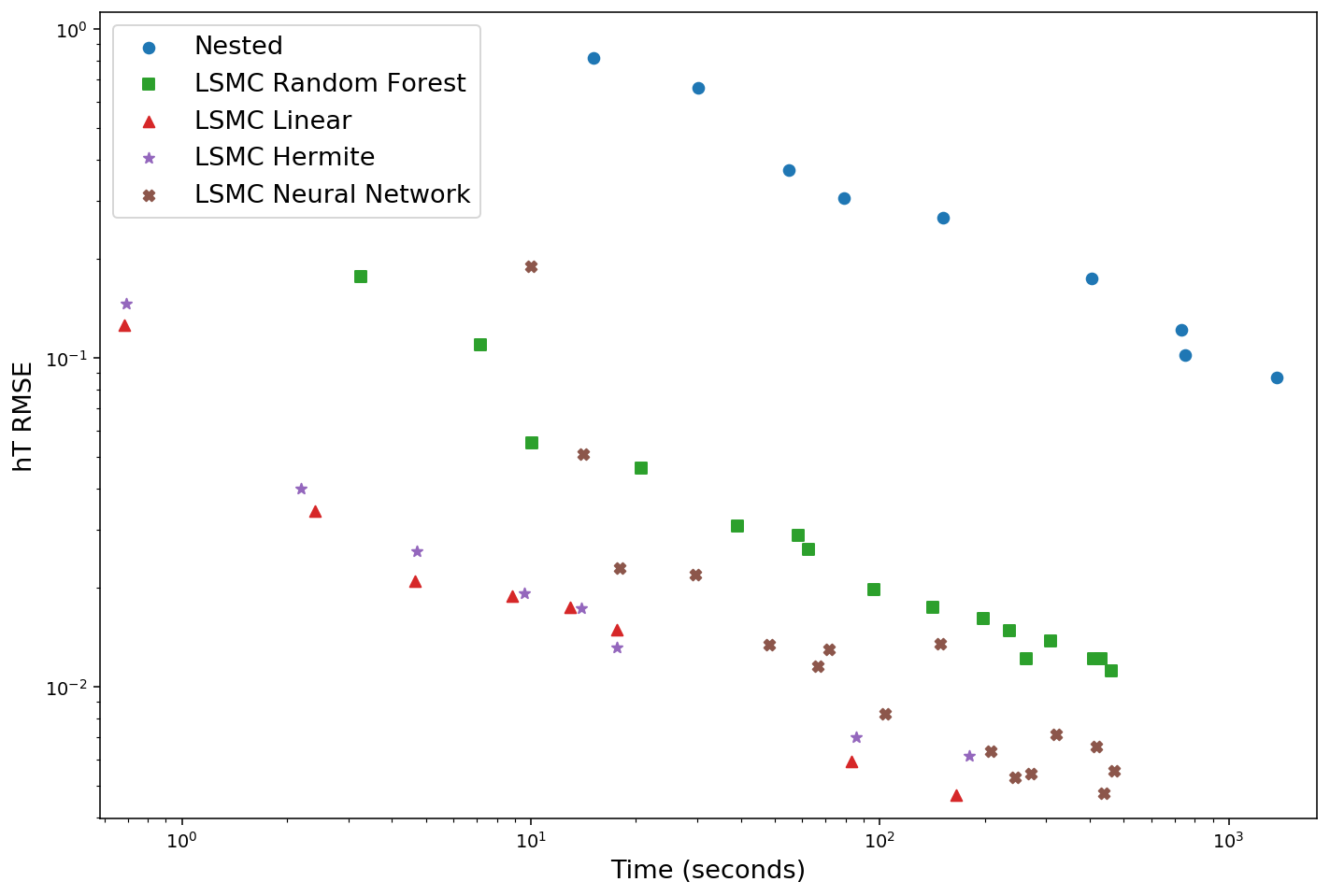}
\else
  \includesvg[width=345 pt]{hT-inner-ode}
\fi
\caption{Evolution of the RMSE of $h_T$ with running time}
\label{fig:hT-mse-ode}
\end{figure}

In \cref{fig:hT-mse-ode}, we observe the evolution of the root mean squared Error of $h_T$. As could be inferred from \cref{hT-compare-ode}, the NMC presents much larger errors for the same computing cost. In all LSMC variants, the error quickly diminishes as the number of paths increases.

\begin{figure}
\centering
\ifdefined\UsePng
  \includegraphics[width=345 pt]{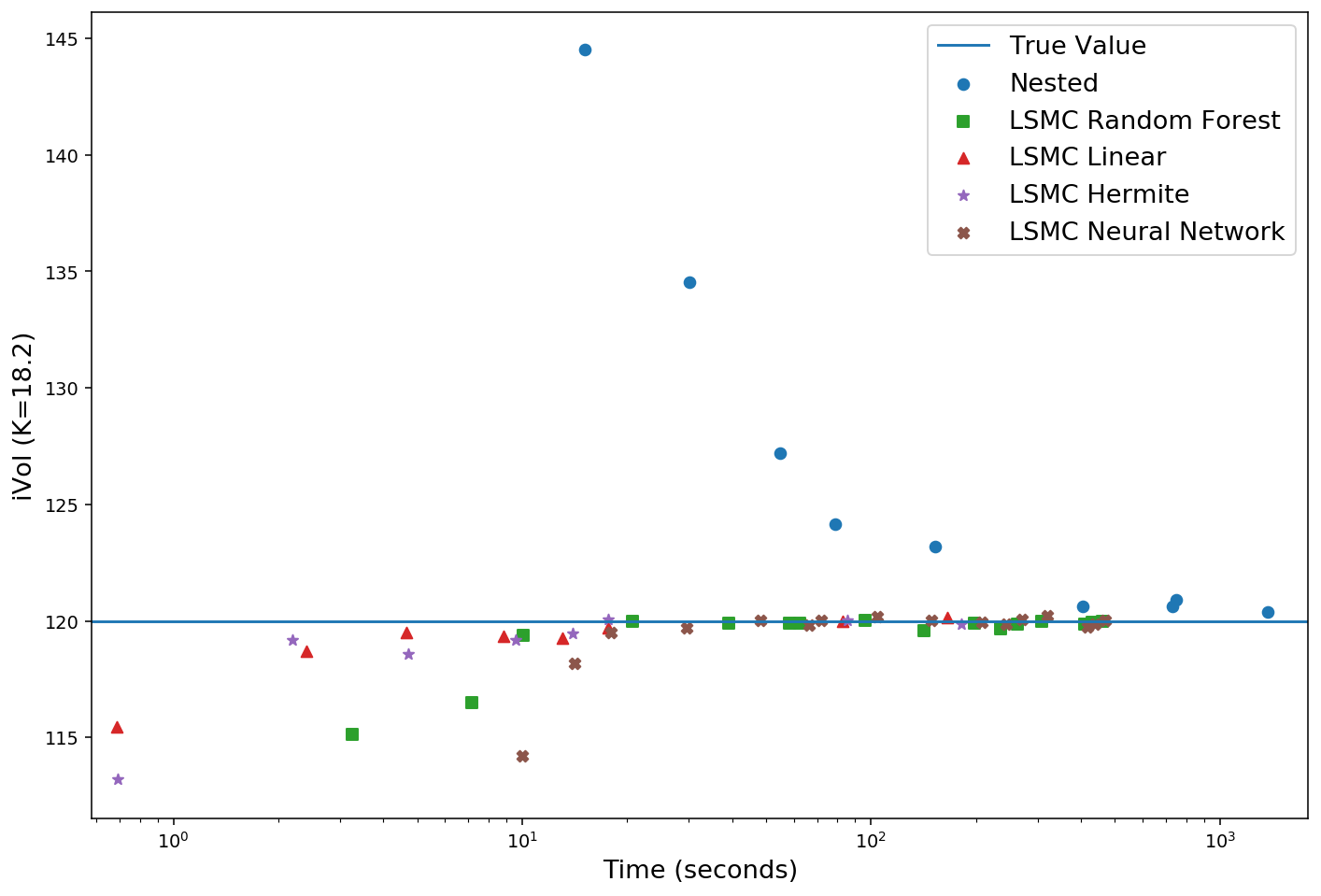}
\else
  \includesvg[width=345 pt]{iVol-inner-ode}
\fi
\caption{Convergence of implied volatility}
\label{fig:iVol-inner-ode}
\end{figure}

\begin{figure}
\centering
\ifdefined\UsePng
  \includegraphics[width=345 pt]{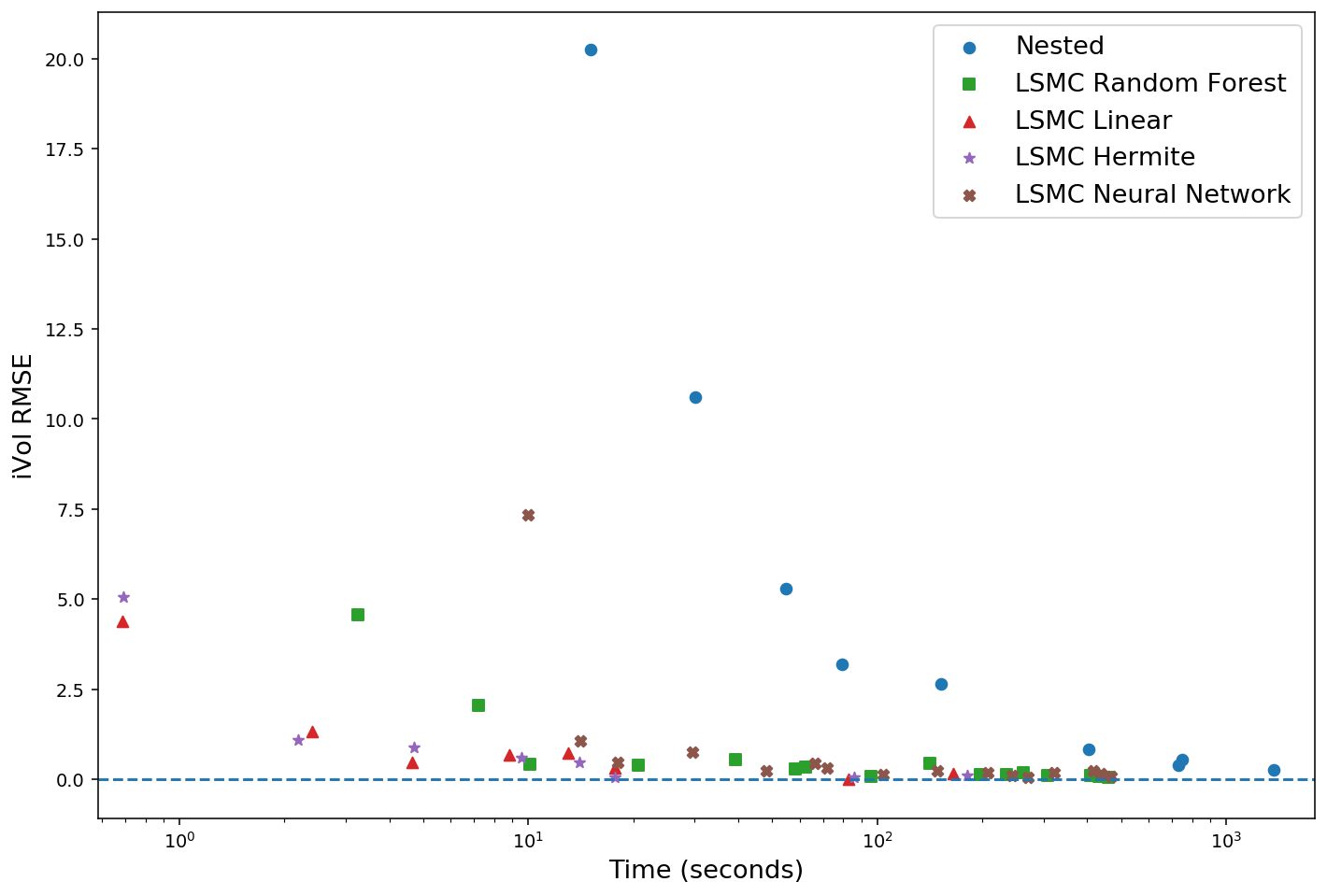}
\else
  \includesvg[width=345 pt]{iVol-mse-inner-ode}
\fi
\caption{Evolution of the RMSE of implied volatility with running time}
\label{fig:iVol-mse-inner-ode}
\end{figure}

The variance present in the estimates of $h_T$ for the NMC will increase the VIX variance and consequently the price of VIX options. For this reason, we observe a consistent positive bias in the NMC estimates for VIX options and implied volatilities. This is apparent in \cref{fig:iVol-inner-ode}. Again, even the more general LSMC variants exhibit fast convergence to the true value. The behavior is similar in \cref{fig:iVol-mse-inner-ode}, where we can observe the evolution of the RMSE with respect to the entire implied volatility smile.

\subsubsection{Non-linear data}
\label{subsec:non-linear}
\begin{figure}
\centering
\includegraphics[width=345 pt]{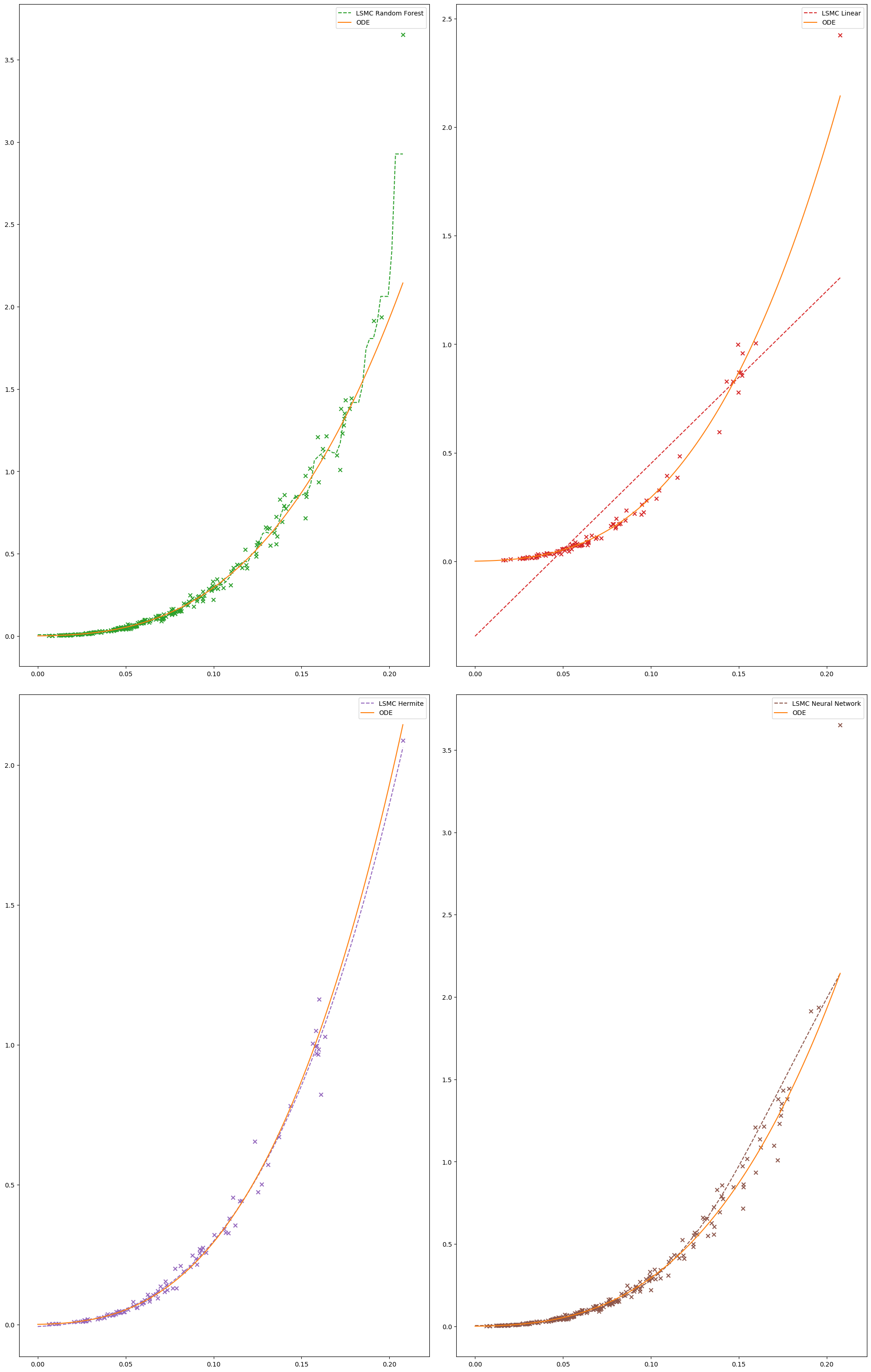}
\caption{Estimates for $h_T$ for various LSMC variants -- non-linear case}
\label{fig:hT-compare-alt}
\end{figure}

As noted before, the LSMC Hermite and LSMC Linear methods are privileged because the true function $h_T$ is log-linear in $\Gamma_T$. To test how these methods perform when the function to learn is not log-linear, we learn $(\log h_T)^3$ instead of $\log h_T$. In \cref{fig:hT-compare-alt}, we can see the behavior of each LSMC in this case. The Random Forest, because of its immense flexibility ends up being sensitive to outliers in sparsely sampled regions. Nevertheless, this does not compromise the learning in more densely sampled regions, which have more importance.  

\begin{figure}
\centering
\ifdefined\UsePng
  \includegraphics[width=345 pt]{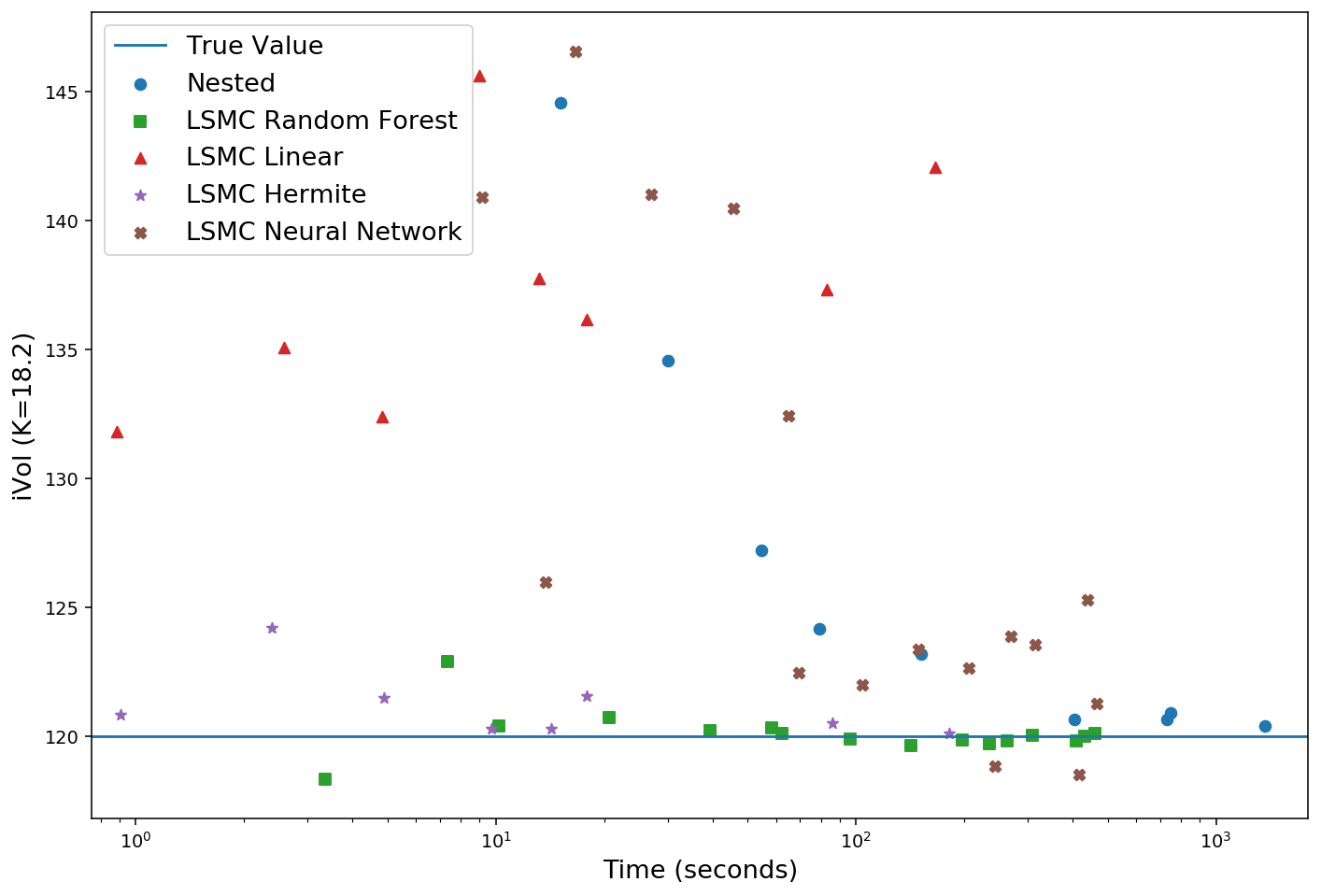}
\else
  \includesvg[width=345 pt]{iVol-inner-sqrt-ode}
\fi
\caption{Convergence of implied volatility -- non-linear case}
\label{fig:iVol-inner-sqrt-ode}
\end{figure}

In \cref{fig:iVol-inner-sqrt-ode}, we can more accurately access the performance of each LSMC variant in this non-linear setting. The linear regression obviously does not (can not) capture the behavior of $h_T$ and thus does not converge to the true value. Both the LSMC Random Forest and the LSMC Hermite still provide reasonably accurate estimates, indicating their reliability even when the true structure of the mapping $h_T$ is unknown.

The LSMC Neural Network exhibits some variance in its estimates and a slower convergence. Perhaps it is possible to tailor the Neural Network architecture to the problem at hand and improve its performance. However, optimizing Neural Network architecture is outside the scope and goal of this paper. Because Neural Networks are more complex, hyper-parameter optimization is thus harder when compared to other variants of the LSMC.

\subsubsection{Performance metrics}

\begin{figure}
\centering
\ifdefined\UsePng
  \includegraphics[width=345 pt]{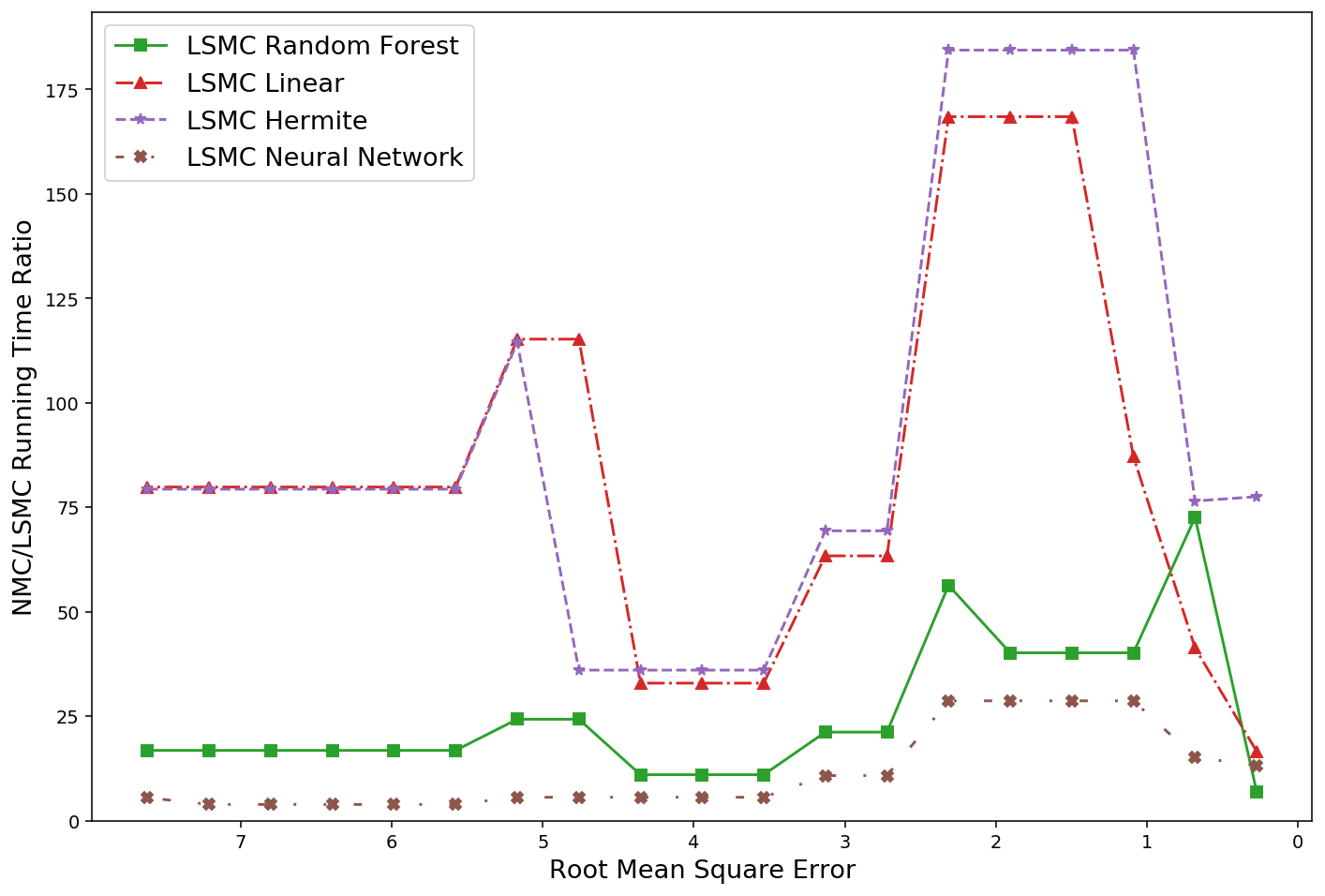}
\else
  \includesvg[width=345 pt]{metric-ratio-time-ode}
\fi
\caption{Relative comparison of running times}
\label{fig:ratio}
\end{figure}

If we wish to compare LSMC methods with the NMC, we can compare their running times. In \cref{fig:ratio}, we compute the ratio of the NMC running time and each LSMC running time for a given goal of iVol RMSE. Because the LSMC Random Forest and LSMC Neural Network have time consuming regressors and require higher values of $N$, their performance is not as high as the LSMC Hermite. Nevertheless, they still outperform the NMC by a considerable factor. 
The Neural Network factor ranges from 3.88 to 28.67, while the Random Forest ranges from 6.93 to 72.57. The LSMC Hermite is the best performing method, preserving adaptability and outperforming the NMC by a huge factor.

\begin{figure}
\centering
\ifdefined\UsePng
  \includegraphics[width=345 pt]{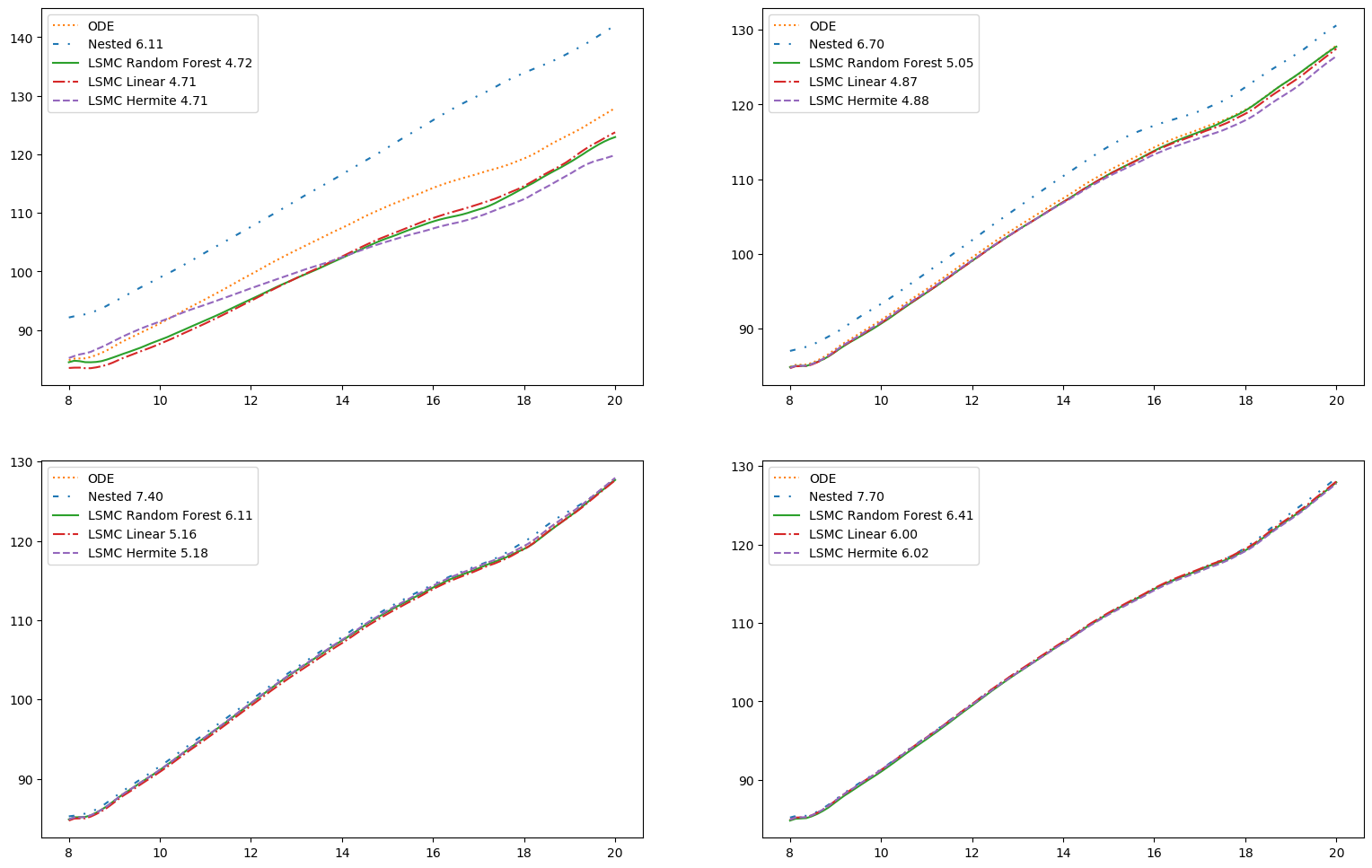}
\else
  \includesvg[width=345 pt]{smile-ode}
\fi
\caption{Comparison of produced iVol Smiles}
\label{fig:smile}
\end{figure}

Finally, in \cref{fig:smile}, we can observe the different volatility smiles produced by the different methods for different number of generated paths. We present the base 10 logarithm of the total number of paths  $N \times K$ in the labels.  

We may thus conclude the NMC, although valid, is very computationally expensive, and is deemed impractical in a production environment. On the other hand, the LSMC method, across its different variants, performs well in pricing VIX derivatives, being able to produce accurate results by a fraction of the computational cost of the NMC. In particular, the LSMC Hermite, frequently deployed in a LSMC setting, shows both fast computing time and adequate robustness. 

\FloatBarrier
\subsection{Dependent vol-of-vol}

Let us now consider the case where the vol-of-vol process $\Gamma$ is no longer independent of $W$ but still Markovian. In this case, the method presented in \cite{ModulatedVolterra} is no longer available, so we cannot compute metrics such as the RMSE. This is, of course, where the LSMC shines.

Although now $h_T(u)$ is not necessarily log-linear in $\Gamma_T$, we will still consider $\log h_T$ as the function to predict and include the simple linear regression model in our set of LSMC methods. This lets us examine how well $h_T$ resembles a log-linear function as in the previous case. If $h_T$ happens to still (even approximately) keep the log-linear property, the simple linear regression will be a useful LSMC method. It is worth noting that if were to drop the Markovianity assumption and $\log h_T$ was a linear functional of the infinite dimensional object that encodes the information at $\mathcal{F}_T$, the VIX pricing problem via LSMC would be reduced to fitting a high dimensional linear regression.
In order to model dependence, we simply let the sBm $(B, W, Z)$ form a 3-dimensional sBm with correlation  matrix
$$
\Sigma = \begin{bmatrix}
1 & \rho & \rho_S \\
\rho & 1  & \rho_V \\
\rho_S & \rho_V & 1 \\
\end{bmatrix}.
$$
We set $\rho_S = -0.9$ and $\rho_V = 0.9$ so that there is strong dependence. The rest of the parameters remain as in the previous section.

\subsubsection{Exploring the data}
\begin{figure}[H]
\centering
\includegraphics[height=500 pt, width=345 pt]{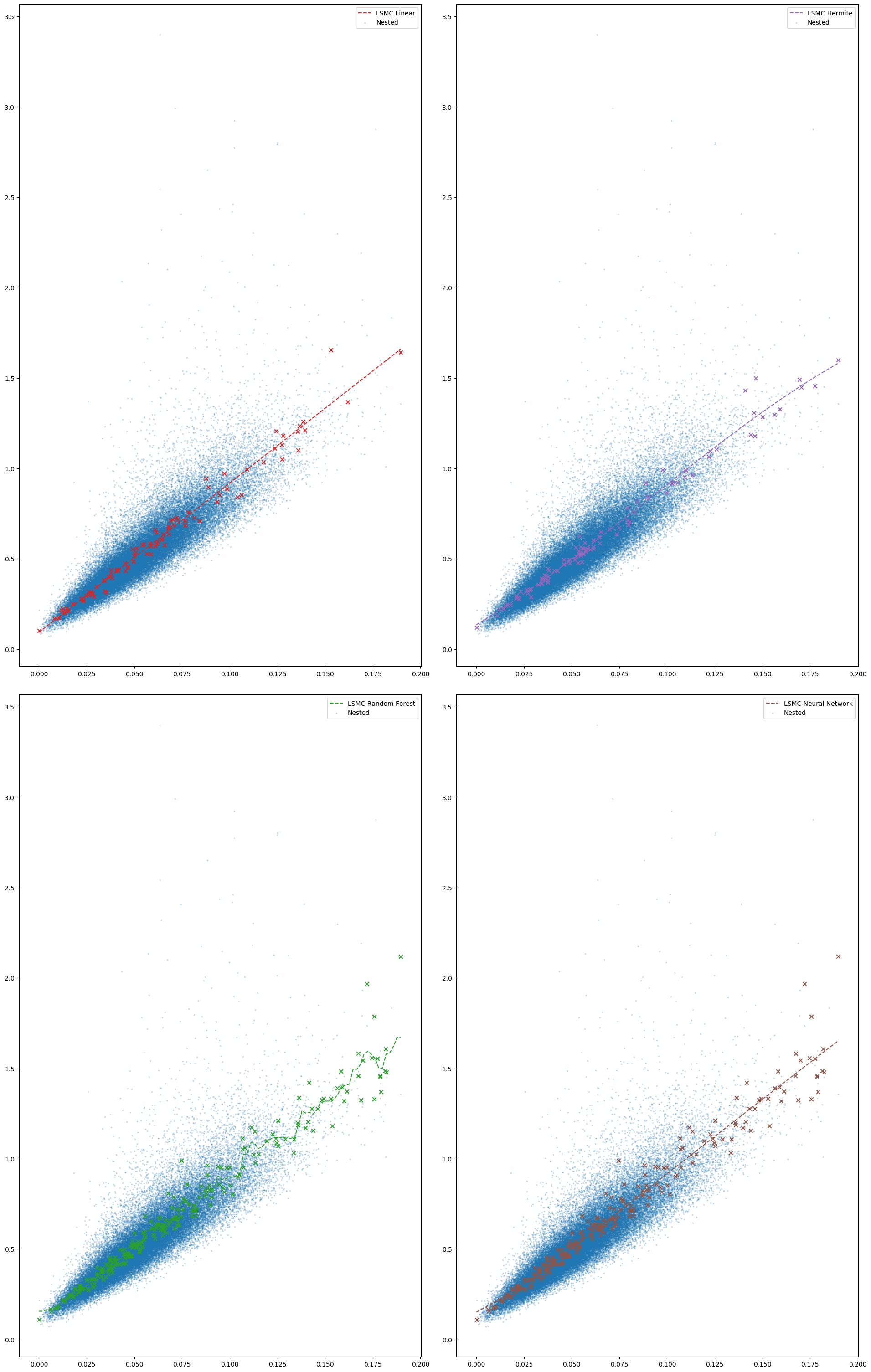}
\caption{Estimates of $h_T$ for various LSMC variants -- dependent $\Gamma$}
\label{hT-compare-dep}
\end{figure}

When compared to the previous case, there is somewhat higher variance because of the high correlation, as can be observed in \cref{hT-compare-dep}. 

\newpage
\FloatBarrier
\subsubsection{Convergence of implied volatility}
We are not in possession of the true value of $h_T$ for a given outer path, but given the results in the previous section, we expect the LSMC methods to converge to the true value. It is not guaranteed the simple linear regression LSMC will converge to the same value, since there is no guarantee that $h_T$ is log-linear in $\Gamma_T$, but from the data exploration conducted previously, it looks it might at least be well approximated by a log-linear function. 

\begin{figure}[H]
\centering
\ifdefined\UsePng
  \includegraphics[width=345 pt]{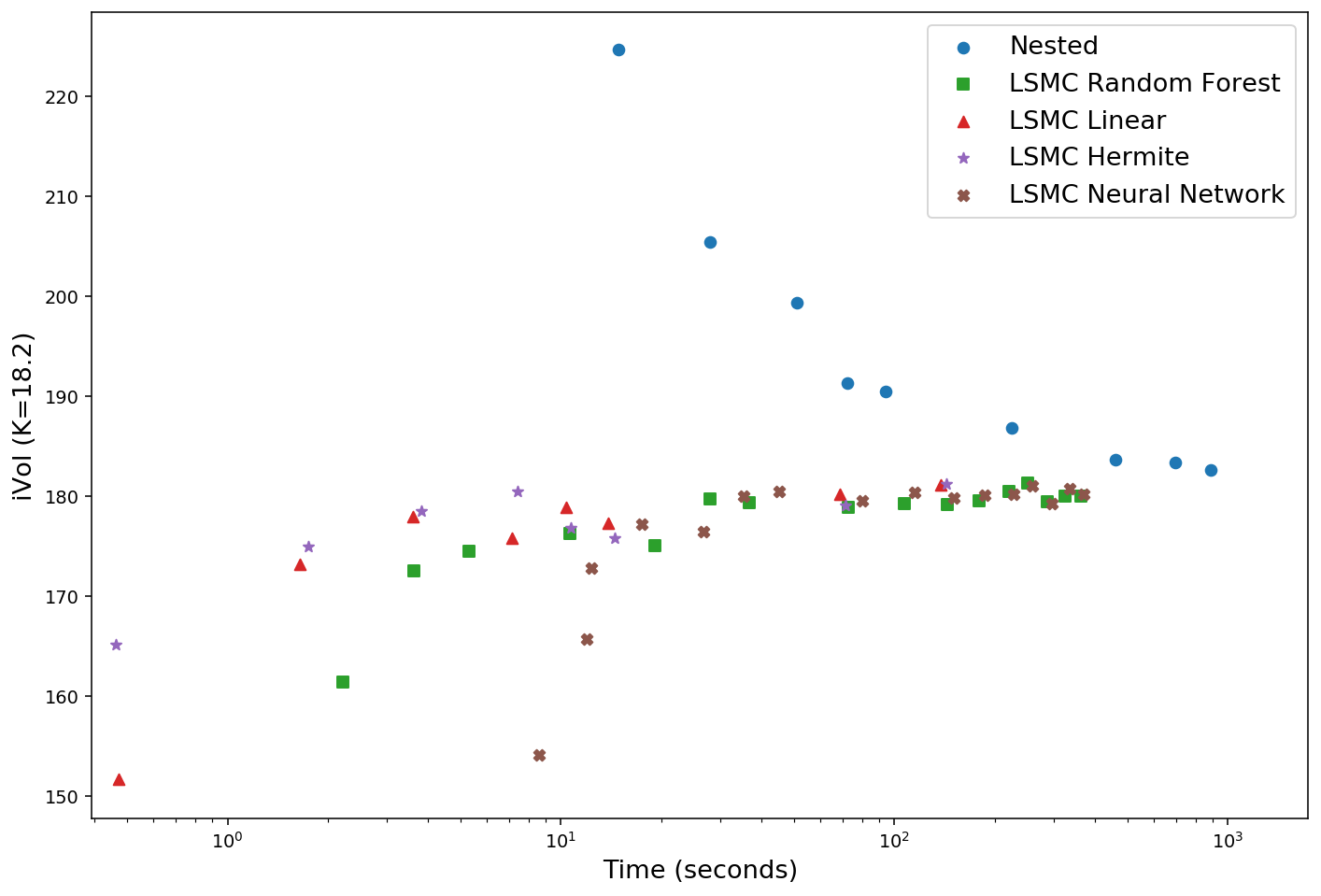}
\else
  \includesvg[width=345 pt]{iVol-inner-dep}
\fi
\caption{Convergence of implied volatility -- dependent $\Gamma$}
\label{fig:iVol-inner-dep}
\end{figure}

Because the data exhibits more variance, we expect slower convergence and a higher bias of the NMC when estimating implied volatility. Nevertheless, we expect both the NMC and the LSMC to converge towards the same value. This behavior is confirmed in \cref{fig:iVol-inner-dep}.

\newpage
\subsubsection{Generated smile}

\begin{figure}[H]
\centering
\ifdefined\UsePng
  \includegraphics[width=345 pt]{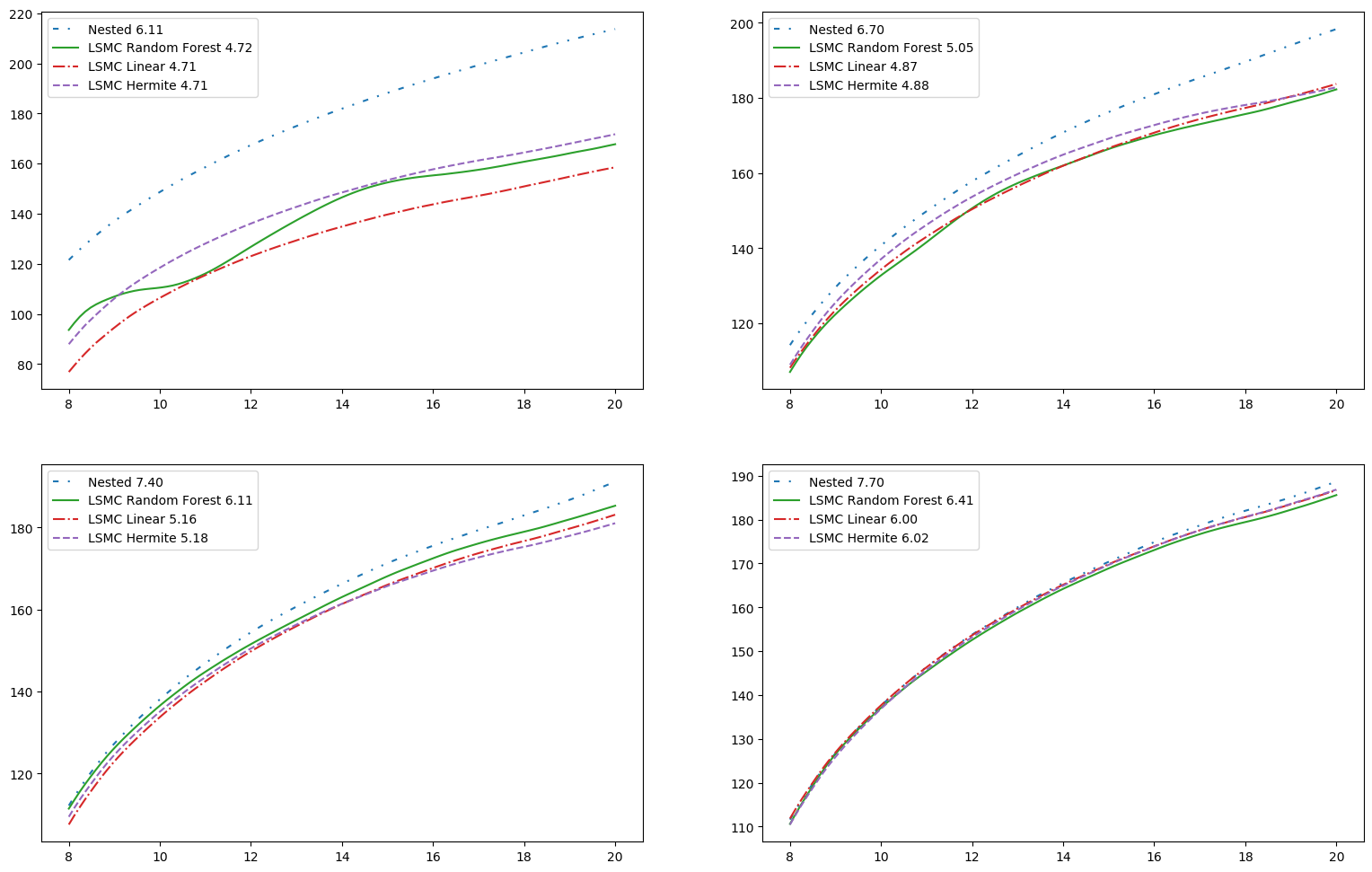}
\else
  \includesvg[width=345 pt]{smile-dep}
\fi
\caption{Comparison of produced implied volatility smiles -- dependent $\Gamma$}
\label{fig:smile-dep}
\end{figure}

In \cref{fig:smile-dep}, we can see the smiles generated in the dependent case, which we may compare to \cref{fig:smile}, to see the impact of adding correlation. Although slowly than in \cref{sec-IndMark}, the generated smiles approach a consistent limit as the computational budget increases.

\clearpage
\FloatBarrier
\section{Conclusion and Further Research} 
\label{sec:conc}

In this paper, we proposed a new method for VIX derivatives pricing in stochastic Volterra models. This method works even in the more generalized framework of SVM, where to the best of our knowledge no VIX pricing method had been proposed. We attested to the good performance of our method using various numerical simulations.

The least squares Monte Carlo method presented and discussed in this paper provides a viable alternative to price VIX derivatives in the context of a generalized framework of stochastic Volterra models. Compared to the nested Monte Carlo method, it requires much lesser computation cost to achieve the same error rate.

The choice of regression model plays a very important role in the LSMC. Ideally, it should be general enough to not require knowledge of the structure of $h_T$. Moreover, it should be robust to outliers, since estimates of $h_T$ are potentially noisy. Finally, it needs to be itself not too computationally expensive, both in terms of memory and computing time. The LSMC Hermite has these characteristics but does not generalize well to higher dimensions. We explored how Random Forests and Neural Networks can handle the one-dimensional case, but more progress can be done with respect to optimizing these kinds of models. Thus, a robust and flexible method more tailored to stochastic Volterra models is a potential area of research. In particular, it would be interesting to develop a full non-Markovian method capable of efficiently handling the rough vol-of-vol case.

Finally, being able to calibrate these models is essential for their usage in industry. Directly applying a conventional calibration algorithm to the LSMC method here presented as it is, may be impractical. However, using the Deep Calibration approach of \cite{DeepBlanka}, a dataset of VIX option valuations can be generated via the LSMC, and then a Neural Network can be trained to learn the mapping. This Neural Network can then be fed to a classical calibration algorithm like the Levenberg-Marquardt. The design of such network and posterior model calibration is also a problem to be studied in further research.

\clearpage
\appendix 
\section{Appendix - Markovian Independent vol-of-vol}
\label{App1}

In this section we present an important result proved in \cite{ModulatedVolterra} that allows us to calculate the infinite dimensional state variable $h_T$, defined in \cref{eq:hT}.
We also provide a proof of \cref{prp:infdim}.

In the model of \cite{ModulatedVolterra}, the vol-of-vol $\Gamma$ is assumed conservative, time-homogeneous and affine. By Theorem 2.7 and Proposition 9.1 of \cite{ModulatedVolterra}, the infinitesimal generator of $\Gamma$ can be written as 
$$
\mathcal{L} f(x) = k (\theta - x) \frac{ \partial f}{\partial x} (x) + \frac{\delta^2 x}{2} \frac{\partial^2 f}{\partial x^2} (x)  + \int_0^{+\infty} [f(x+z)-f(x)][m(dz)+x\mu(dz)]
$$
for constants $k, \theta, \delta \geq 0$ and $m, \mu$ positive measures on $\R^+$ such that 
$$
\int_0^{+\infty} (z \land 1)[ \mu(dz) + m(dz)] <  +\infty.
$$
\begin{thm} [adapted from Prop. 3 in \cite{ModulatedVolterra}]
\label{thm:ModVol}
Let $T>0$ and define
$$
R(u) =-ku + \frac{\delta^2}{2}u^2 + \int_0^{+\infty} (e^{zu}-1) \mu(dz)
$$
$$
F(u) = k\theta u + \int_0^{+\infty} (e^{zu}-1) m(dz)
$$
Suppose that there exists $A>0$ such that
$$
\int_1^{+\infty} z e^{zA} [ \mu(dz) + m(dz)] < +\infty
$$
$$
2G(T) + T(0 \lor R(A)) \leq A.
$$
Then, the Ricatti ODE 
$$
\frac{\partial}{\partial t} \psi(t) = 2g^2(t) + R(\psi(t))
$$
$$
\frac{\partial}{\partial t} \phi(t) = F(\psi(t))
$$
with initial condition $\psi(0) = \phi(0) = 0$ has a solution on $[0,T]$, where $0 \leq \psi \leq A$ and $0 \leq \phi \leq TF(A)$. Moreover
$$
h_t(u, \gamma) := \ev{ \exp \left( 2 \int_t^u g(u-s)^2 \Gamma_s \, ds \right) \mid \Gamma_t = \gamma }
= \exp\left[ \gamma \psi(u-t) + \phi(u-t) \right].
$$
\end{thm}

Let us now present a simple proof of \cref{prp:infdim}.

\begin{proof}[Proof of \cref{prp:infdim}]
We start by noting that
\begin{align*}
\xi_t(u)
&= \ev{v_u \mid \mathcal{F}_t} \\
& =A_0(u) \ev{ E_{0, t}(u) E_{t, u}(u) \mid \mathcal{F}_t} \\
&= A_0(u) E_{0, t}(u) h_t(u) \\
&=\frac{ \xi_0(u) }{h_0(u) } E_{0, t}(u) h_t(u) .
\end{align*}
Thus
$$
A_t(u) = A_0(u) E_{0, t}(u)
$$
and hence
$$
v_u = A_t(u) E_{t, u}(u).
$$
If $\Gamma$ is assumed Markovian, given $\Gamma_t$, $E_{t, u}$ is independent of $\mathcal{F}_t$.
\end{proof}

\section{Appendix - Random Forests}
\label{App2}
In this section, we present some important results found in \cite{RF}, concerning ensemble methods in general and Random Forests in particular. When trying to learn the function $h_t$ of \cref{eq:hT}, we may not know its structure in advance. Thus, we wish to use a regression model that is capable of learning a large range of functions, which often comes with added complexity. The price to pay for this, unfortunately, is overfitting. The main advantage of ensemble methods is that they allow us to reduce overfitting while at the same time not constraining the class of functions the regression can learn.

The decision tree regressor essentially approximates the target function by a step function. Since step functions are dense in $L^1$, decision trees are able to approximate a wide range of functions. The main issue with decision trees is their propensity to overfit. Overfitting can be loosely defined as an extreme sensitivity to training data. More formally, we can measure it in terms of the variance of model predictions with respect to the (random) data used train the model. In order to make a rigorous treatment of these matters, let us establish some notation.

Let $X$ and $Y$ be random variables in some probability space, taking values in $\X$ and $\Y$, respectively. We wish to give our best prediction of $Y$ given an observed value of $X$. A regression model $\varphi$ is a function such that, given training data $\LL \in \mathcal{P}( \X \times \Y)$, produces a function $\varphi_\LL : \X \to \Y$. To this function we can associate the random variable $\varphi_\LL(X)$. The mean squared error (MSE) of the regression model given the training data $\LL$ can be defined as
$$
MSE( \varphi \mid \LL) = \ev{
\left( \varphi_\LL(X) - Y\right)^2
},
$$
where the expectation is taken over the probability space where $X$ and $Y$ exist but the function $\varphi_\LL$ (and hence the training data) is fixed.  If $Y=f(X)$ is a deterministic function of $X$, the expectation is taken with respect to a deterministic random variable. In this case, it is possible that $\varphi_\LL$ predicts $Y$ perfectly (if $\varphi_\LL = f$). If, on the other hand, $Y$ is not solely determined by $X$, there may be a positive lower bound to $MSE(\varphi \mid \LL, X =x)$. The Bayes regression model is defined as the one that achieves such lower bound, and it does not depend of any training data $\LL$. It is of course given by
$$
\phi_B(x) = \ev{ Y \mid X=x}.
$$
Since we usually do not know the distribution of $X \times Y$, we have no access to the Bayes regressor. Note that the Bayes regressor is optimal for each possible value $x$ of $X$. Indeed, it minimizes the MSE for  each fixed $X=x$:
$$
MSE(\varphi \mid \LL, X =x) = 
\ev{
\left( \varphi_\LL(x) - Y\right)^2 \mid X = x
}.
$$
In the above expectation, only $Y$ is random, and it follows its conditional distribution given $X=x$.

We are interested in knowing the model error across all $\LL$ for each $x \in X$
$$
MSE(\varphi \mid X= x) = \ev{  MSE( \varphi \mid \LL, X = x) }.
$$
Here the expectation follows the distribution of all possible training sets $\LL$ and that of $Y$ conditioned on $X=x$. 

\begin{thm}[Bias Variance Decomposition (thm 4.1. in \cite{RF}]
Let $x \in \X$. Let the MSE for the Bayes model be denoted by
$$
noise(x) = MSE(\varphi_B \mid X= x).
$$
Consider the average prediction of our regression model $\varphi$ across all possible training sets 
$$
\bar{\varphi}(x) := \ev{ \varphi_\LL(x)}
$$
and define the bias with respect to the Bayes model
$$
bias(x) = \varphi_B(x) - \bar{\varphi}(x)  = \ev{Y - \bar{\varphi}(X) \mid X=x}.
$$
Let us also denote the variance of $\bar{\varphi}(x)$ across training sets by
$$
var(x) = \ev{
\left(
\varphi_\LL(x)  - \bar{\varphi}(x)
\right)^2
}.
$$
then the MSE can be decomposed as follows
$$
MSE(\varphi \mid X= x) = noise(x) + bias^2(x) + var(x).
$$
\end{thm}
the term $noise(x)$ is independent of the choice of regression model and can be seen as the intrinsic lower bound of the $MSE$ posed by the problem. The Bayes regressor of course has zero bias and variance and thus attains this lower bound. For our choice of regression model, we must examine the bias-variance trade-off. A simpler model might achieve lower variance but be biased. A complex model might have small or even zero bias but might have high variance and thus be very sensible to training data. This phenomenon is what is usually known in Machine Learning jargon as overfitting.

the main idea of ensemble methods such as Random Forests, is to reduce overfitting, that is $\var(x)$, while not increasing the bias. They are composed of $M$ randomized models and their prediction is given by the average across each member of the ensemble
$$
\varphi(x \mid \LL, \theta) = \frac{1}{M}\sum_{i=1}^M \varphi^i_\LL(x),
$$
for randomized models $\varphi^i$  (ensemble members). In order to be rigorous, let us say each randomized regressor $\varphi^i$ is determined by the value of certain random parameter $\theta_i$. Note that now the trained regression model depends not only on the choice of the training set $\LL$ but also of the choice of $\theta_i$.  thus, the expected value in the mean prediction for the ensemble method is taken across both training sets and random parameters $\theta$
$$
\bar{\varphi}(x) = \ev{ \ev{ \varphi_{\LL,\theta} (x)} }.
$$
For a fixed training set $\LL$ and a fixed $x$, $\varphi_i(x)$ is a random variable, where the randomness exists exclusively through $\theta_i$.
\begin{thm}[thm 4.3. in \cite{RF}]
Assume the $\theta_i$ are i.i.d.. Thus, the variance is equal across each member of the ensemble:
$$
\sigma^2(x) = var_{\varphi^i}(x) \, \forall \, i = 1, ..., M.
$$
By linearity of the expectation and the tower property, the bias of the ensemble is the same as the bias of any individual member
$$
bias(x) = \varphi_B(x) -\bar{\varphi}(x) =  \varphi_B(x) -\bar{\varphi^i}(x).
$$
Let $\rho(x)$ denote the correlation coefficient between $\varphi^i(x)$ and $\varphi^j(x)$. These two random variables only depend on $\theta_i$ and $\theta_j$, which are assumed i.i.d.. Thus, $\rho(x)$ does not depend on $i$ and $j$. Then $\rho(x) \geq 0$ and the variance of the ensemble is given by 
$$
var(x) = \sigma^2(x) \left( 
\rho(x)  + \frac{1-\rho(x)}{M} 
\right).
$$
In particular, $var(x)  \leq \sigma^2(x)$. 
\end{thm}
Thus, the stronger the random effects in the ensemble (meaning $\rho(x) \to 0$), the greater is the variance reduction and hence the lower the MSE of the ensemble method.

We conclude by noting that Random Forests do not totally solve the problem of overfitting but are an improvement when compared to decision trees. The improvement is more evident in higher dimensional problems with many predictors, since there it is possible to randomize the predictors across each different tree in the forest. This should inform our analysis of the Random Forest performance in lower dimensional problems such as the one presented in this paper.

\clearpage
\bibliography{bib}

\end{document}